%Paper: hep-th/9503017
%From: skyang@het.ph.tsukuba.ac.jp (Sung-Kil Yang)
%Date: Fri, 3 Mar 95 15:14:54 JST
%Date (revised): Fri, 3 Mar 95 15:45:26 JST

\input phyzzx

\rightline{hep-th/9503017, UT-700}
\date{March, 1995}
\titlepage
\vskip 1cm
\title{Topological $\sigma$-Models and Large-$N$ Matrix Integral}
\author {Tohru Eguchi and Kentaro Hori}
\address{Department of Physics, Faculty of Science, University of
Tokyo, Tokyo 113, Japan}
\author{Sung-Kil Yang}
\address{Institute of Physics, University of Tsukuba,
Ibaraki 305, Japan}

\abstract{
In this paper we describe in some detail the representation of
the topological $CP^1$ model in terms of a matrix integral
which we have introduced in a previous article. We first discuss the
integrable structure of the $CP^1$ model and show
that it is governed by an extension
of the 1-dimensional Toda hierarchy. We then introduce a matrix model
which reproduces the sum over holomorphic maps from arbitrary Riemann
surfaces onto $CP^1$. We compute intersection numbers on the moduli space
of curves using geometrical method and show that the results
agree with those predicted by the matrix model. We also develop a
Landau-Ginzburg (LG) description of the $CP^1$ model using a superpotential
$e^X+e^{t_{0,Q}}e^{-X}$ given by the Lax operator of the Toda hierarchy
($X$ is the LG field and $t_{0,Q}$ is the coupling constant
of the K\"ahler class). The form of the superpotential indicates the close
connection between $CP^1$ and $N=2$ supersymmetric sine-Gordon theory which
was noted some time ago by several authors.
We also discuss possible generalizations of our construction to other
manifolds and present a LG formulation of the topological $CP^2$ model.}

\endpage
\overfullrule=0pt

%%%%%%%%%%%%%%%%%%%%%%% definitions  %%%%%%%%%%%%%%%%%%%%%%
\def\cmp#1{Commun. Math. Phys. {\bf #1}}
\def\pl#1{Phys. Lett. {\bf B#1}}

\def\np#1{Nucl. Phys. {\bf B#1}}
\def\ijmp#1{Int. J. Mod. Phys. {\bf A#1}}
\def\mpl#1{ Mod. Phys. Lett. {\bf A#1}}
\def\e{\hfill\break}

\def\Desc#1#2{\sigma_#1(#2)}
\def\half{{1 \over 2}}
\def\S{\sigma}
\def\C{t_{0,P}}
\def\T{t_{0,Q}}
\def\bra{\langle}
\def\ket{\rangle}
\def\PP{\bra PP \ket}
\def\PQ{\bra PQ \ket}
\def\pder#1{{\partial \over \partial t_{#1}}}
\def\longto{\longrightarrow}
\def\CP{{\bf C}{\rm P}^1}
\def\bfC{{\bf C}}
\def\Md{{\cal M}_{0,1}(\CP,d)}
\def\bMd{\overline{\cal M}_{0,1}(\CP,d)}
%%%%%%%%%%%%%%%%%%%%%%%%%%%%%%%%%%%%%%%%%%%%%%%%%%%%%%%%%

\chapter{Introduction}

In our attempts at understanding the geometrical principles behind the string
theory, the approach of topological field theories
\REF\Wtop{E. Witten, \cmp{117} (1988) 353; \cmp{118} (1988) 411.}
[\Wtop]
seems to offer an important
clue. String theory exposes its geometrical structures in a most transparent
manner in its topological formulation. Thus we may gain geometrical
insights from the study of the topological version of string theories.
When a string is compactified on a K\"ahler manifold $M$, the theory is
described by an $N=2$ supersymmertric non-linear $\sigma$-model.
After suitable twistings an $N=2$ $\sigma$-model
yields a pair of topological field theories, topological
A- and B-models. B-models are essentially classical and easy to solve.
A-model, on the other hand, is given by a sum over the
holomorphic maps (instantons) from the Riemann surface
to the target manifold and in general difficult to evaluate.

When the target space $M$ is a Calabi-Yau manifold, a special situation
arises; A-model may be replaced by the B-model associated with the mirror
manifold of $M$ and the genus=$0$ A-model correlation functions are
obtained from those of the B-model using the mirror map
\REF\Can{P. Candelas, X. de la Ossa, P. Green and L. Parkes,
\np{359} (1991) 21.}
\REF\Oth{
A. Klemm and S. Theisen, \np{389} (1993) 153; \e
A. Font, \np{391} (1993) 358; \e
P. Candelas, X. de la Ossa, A. Font, S. Katz and D. Morrison,
\np{416} (1994) 481;  \e
P. Candelas, A. Font, S. Katz and D. Morrison, \np{429} (1994) 626 \e
S. Hosono, A. Klemm, S. Theisen and S.-T. Yau, \np{433} (1995) 501.}
\REF\higher{
B. Greene, R. Plesser and D. Morrison, ``Mirror manifolds in higher
dimension'', hep-th/9402119; \e
M. Jinzenji and M. Nagura, ``Mirror symmetry and an exact calculation of
$N-2$ point correlation function on Calabi-Yau manifold embedded in
$CP^{N-1}$'',UT-680, hep-th/9409029.}
[\Can,\Oth,\higher].
Recently the A-model partition functions are determined further at higher
genera $g=1,2$ using the method of holomorphic anomaly
\REF\BCOV{M. Bershadsky, S. Cecotti, H. Ooguri and C. Vafa,
\np{405} (1993) 279; \cmp{165} (1994) 311.}
[\BCOV].

When we come to target manifolds $M$ with a positive 1st Chern class $c_1(M)>0$
(Fano varieties), the situation becomes quite different from that of
Calabi-Yau
manifolds which are characterized by the condition $c_1(M)=0$ and
scale invariance. It is well-known that the one-loop $\beta$-function of
a supersymmetric non-linear $\sigma$-model is proportional to the
1st Chern class of $M$. When $c_1(M)>0$, the theory is
asymptotically free and
develops a mass gap due to the dimensional transmutation.
Examples of manifolds with $c_1(M)>0$ are given by
complex projective spaces $CP^n$ and Grassmannians.

Recently an algorithm based on the associativity of the
operator-product-expansion has been used to recursively
determine the instanton sum in the
genus $g=0$ free energy of the A-model with the target manifolds $CP^2,CP^3,
\cdots$ etc.
\REF\KM{M. Kontsevich and Yu. Manin, \cmp{164} (1994) 525.}
\REF\D{B. Dubrovin, ``Geometry of 2D topological field theories'',
SISSA-89/94/FM,\e  hep-th/9407018.}
\REF\I{C. Itzykson, ``Counting rational curves on rational surfaces'',
Saclay T94/001.}
\REF\DI{P. Di Francesco and C. Itzykson, ``Quantum intersection rings'',
SPhT94/111, hep-th/9412175.}
\REF\JS{M. Jinzenji and Y. Sun, ``Calculation of Gromov-Witten invariants
for $CP^3$, $CP^4$, and Gr(2,4)'', UT-696, hep-th/941226.}
[\KM,\D,\I,\DI,\JS].
This method is efficient and general, however, is limited to the genus $g=0$
case at the moment.

In a previous paper
\REF\EY{T. Eguchi and S.-K. Yang, \mpl{9} (1994) 2893.}
[\EY]
we have considered $CP^1$ , the simplest manifold with
$c_1(M)>0$, and constructed a matrix model which reproduces
holomorphic maps from Riemann surfaces of arbitrary genera onto $CP^1$.
A characteristic feature of the $CP^1$ model is the presence of the logarithmic
terms in the action which implies the scaling violation in the system. In this
paper we would like to describe in some details the properties of the model
and its possible extensions. In section 2 we first briefly recall the basic
facts
about the topological $CP^1$ model and steps in deriving the matrix action.
In section 3 we obtain a set of Ward identities of the model which form a
Virasoro algebra. We check these identities
against the geometrical data of the moduli
space of holomorphic maps
in the case of $g=0,1$. Intersection numbers on the moduli
space are calculated using
the method of algebraic geometry and we find a complete agreement with the
predictions of the $CP^1$ model.
In section 4 we discuss a Landau-Ginzburg (LG)
description of the $CP^1$ model. It turns out that all the $g=0$ correlation
functions are reproduced by residue integrals if we use a superpotential
of the form $\cos X$ where $X$ is the LG field. Thus the topological $CP^1$
model
may be identified with the (topological version of the) $N=2$ supersymmetric
sine-Gordon theory. The close relation between the $CP^1$ and $N=2$ sine-Gordon
models has been noted sometime ago based
on the comparison of the particle spectra
and $S$-matrices of these theories
\REF\CV{S. Cecotti and C. Vafa, \cmp{158} (1993) 569.}
\REF\FI{P. Fendley and K. Intriligator, \np{380} (1992) 265.}
[\CV,\FI].
In Section 5 we discuss a possible generalization of our construction to
other manifolds and formulate a LG description of the topological
$CP^2$ model. We also present comments and discussions.

\chapter{The $CP^1$ model}

Physical observables in the topological A-model arise from the de Rham classes
of the target manifold. In the case of $CP^1$, there are just two classes,
1 (identity) and $\omega$ (K\"ahler class), and the corresponding physical
observables are denoted as $P$ and $Q$, respectively. The
integrable structure of the
system is described using 2-point functions
$$
\PP  =u, \hskip7mm  \PQ =v.
\eqn\uv
$$
The genus zero free energy in the small phase space is given by
$$
F_0=\half \C^2 \T+e^{\T},
\eqn\zerofree
$$
where the parameters $\C$, $\T$ are coupled to $P$, $Q$. The second term in the
RHS of \zerofree\ comes from the contribution of the degree-1 instanton.
Combining \uv\ and \zerofree\ we note that
$$
\bra QQ \ket =e^{\T}=e^{\PP}
\eqn\consti
$$
holds in the small phase space. Using the topological recursion relation it is
possible to show that \consti\ in fact holds in a large phase space where
couplings $\{ t_{n,P},t_{n,Q}, n=1,2,\cdots \}$ to the
gravitational descendants
of $P$, $Q$ do not vanish
\REF\DW{R. Dijkgraaf and E. Witten, \np{342} (1990) 486.}
[\DW].
It is then easy to see that the
integrable structure of the $CP^1$ model is described by the 1-dimensional
Toda hierarchy
$$
{\partial^2 u \over \partial \T^2}
={\partial \over \partial \T}{\partial \over \partial \C} \PQ
={\partial \over \partial \C} \bra PQQ \ket
={\partial^2 \over \partial \C^2} e^u.
\eqn\toda
$$
\toda\ in fact is the  Toda-lattice equation when we identify $\C$
as the continuum version of the index $n$ of a field  $u_n=u(\C)$.

The Lax formalism of the Toda-lattice hierarchy is well-known. We introduce a
Lax operator (at genus $g=0$ or dispersionless limit, it becomes a number
rather than an operator)
$$
L=p+v+e^u p^{-1}
\eqn\lax
$$
and construct Hamiltonians
$$
H_n^Q=[L^n]_+,  \hskip5mm  n=1,2,\cdots .
\eqn\hamil
$$
($+$ means to take terms with non-negative powers of $p$). \hamil\ generate
flows in the Toda times $t_n$ $(n=1,2,\cdots)$
$$
{\partial L \over \partial t_n}=\{ H_n^Q,L \}, \hskip5mm  n=1,2,\cdots ,
\eqno\eq
$$
where $\{ A,B\}$ denotes the Poisson bracket of the Toda theory defined by
$$
\{A,B\}=p\Big( {\partial A \over \partial p}{\partial B \over \partial \C}
-{\partial B \over \partial p}{\partial A \over \partial \C}\Big).
\eqn\poissonbra
$$
It is easy to identify the Toda times $t_n$ as the descendant times
$t_{n-1,Q}\ n=1,2,\cdots$ of the operator $Q$.
For instance, putting $n=1, \phi_\alpha=Q$, and $X=Y=P$ or $X=P$, $Y=Q$
in Witten's topological recursion relation
\REF\Wgrav{E. Witten, \np{340} (1990) 281.}
[\Wgrav]
$$
\bra \S_n(\phi_\alpha)XY \ket =n\bra \S_{n-1}(\phi_\alpha) \phi_\beta \ket
\bra \phi^\beta XY \ket
\eqn\toprec
$$
we obtain flow equations
$$
\eqalign{
{\partial u \over \partial t_{1,Q}}
&=\bra \S_1(Q)PP \ket =\Big( \half v^2+e^u \Big)',  \cr
{\partial v \over \partial t_{1,Q}}
&=\bra \S_1(Q)PQ \ket =(ve^u)'. \cr}
\eqn\spp
$$
Comparing \spp\ with
$$
{\partial L\over \partial t_2}=2(ve^u)'+(v^2+2u)'e^u p^{-1},
\eqno\eq
$$
we find $t_2=t_{1,Q}/2$. In general a relation
$$
t_n={1 \over n}t_{n-1,Q} \hskip1mm , \hskip5mm n=1,2,\cdots
\eqno\eq
$$
holds.

Analysis of the flows in the parameters $\{ t_{n,P}, n=1,2,\cdots \}$ is
more involved.
Flow equations can again be written down using the recursion relation
\toprec. We have, for instance,
$$
\eqalign{
{\partial u \over \partial t_{1,P}}
&=\bra \S_1(P)PP \ket =(uv)',  \cr
{\partial v \over \partial t_{1,P}}
&=\bra \S_1(P)PQ \ket =\Big(\half v^2+(u-1)e^u \Big)'. \cr}
\eqn\pflow
$$
It is somewhat non-trivial to find Hamiltonians for these flows.
It turns out [\EY] that Hamiltonians involving logarithms of the Lax
operator
$$
\eqalign{
H_n^P=2 & [L^n(\log L-c_n)]_+, \hskip5mm n=0,1,\cdots , \cr
& c_n=\sum_{j=1}^n 1/j, \hskip5mm c_0=0 \cr}
\eqn\phamil
$$
generate the flows in the $t_{n,P}$ variables. In \phamil\ the logarithm
of $L$ is defined by taking the average
$$
\eqalign{
\log L &=\log (p+v+e^up^{-1})  \cr
&=\half \log p(1+vp^{-1}+e^up^{-2})
+\half \log e^up^{-1}(1+ve^{-u}p+e^{-u}p^2)  \cr
&={u \over 2} +\half \log (1+vp^{-1}+e^up^{-2})
+\half \log (1+ve^{-u}p+e^{-u}p^2).  \cr}
\eqno\eq
$$
We note in particular an important relation
$$
{\partial L\over \partial t_{0,P}}
=\{ H_0^P,L\} =2 \{ [\log L]_+,L\}.
\eqno\eq
$$
Note that $\{ H_n^P \}$ mutually commute with each other and also with
$\{ H_n^Q \}$.

So far we have considered the $CP^1$ model at genus $g=0$. Higher genus
structure of the theory can be described by a matrix model where
$L$ becomes a matrix acting on the space
of orthogonal polynomials and Poisson brackets are replaced by matrix
commutators.
Let us recall the system of orthogonal polynomials of a matrix model with an
action $S$,
$$
\int d\lambda \varphi_n(\lambda) \varphi_m(\lambda)e^{NS(\lambda)}
=\delta_{nm}h_n, \ \ n,m=0,1,\cdots,N-1.
\eqno\eq
$$
Here $\varphi_n(\lambda)=\lambda^n+(\hbox{lower order terms})$ are degree-$n$
polynomials which are orthogonal to each other with respect to the weight
$\exp NS(\lambda)$ ($N$ is the genus expansion parameter).
Then the multiplication by $\lambda$ is represented by a matrix $Q$
$$
\lambda \varphi_n(\lambda)=\sum_{m=n-1}^{n+1} Q_{nm}\varphi_m(\lambda)
\eqno\eq
$$
which has non-vanishing elements along 3 diagonal lines $m=n,n\pm 1$.
We parametrize the matrix $Q$ as
$$
Q_{nm}=\delta_{n+1,m}+v_n\delta_{n,m}+e^{N(\phi_n-\phi_{n-1})}\delta_{n-1,m},
\hskip7mm h_n=e^{N \phi_n}.
\eqn\qmatrix
$$
In the continuum limit, $n/N$ is replaced by $\C$ and \qmatrix\ becomes the
dispersionless Lax operator $L$ \lax. Matrix commutators are replaced
by Poisson brackets \poissonbra.

Now the partition function of the $CP^1$ model is given by an matrix integral
[\EY]
$$
Z=\int dM e^{N {\rm Tr} S(M)},
\eqn\partition
$$
where $M$ is an $(N\times t_{0,P})^2$ hermitian matrix and the action $S(M)$ is
defined by
$$
S(M)=-2 M(\log M-1)
+\sum_{n=1}{t_{n-1,Q} \over n}M^n+\sum_{n=1}2t_{n,P}M^n(\log M-c_n).
\eqn\action
$$
Note that the 1st term in the action has the same form as the piece
proportional to $t_{1,P}$ and hence we may consider $t_{1,P}$ having
a ``background value'' $-1$.
The structure of $S(M)$ may be intuitively inferred from the forms of the
Hamiltonians $H_n^Q$ \hamil\ and  $H_n^P$ \phamil.

The characteristic feature of \action\ is the appearance of logarithms which
generate additive terms under the scale transformation of the matrix $M$.
As we see in the next section, this is the mechanism by means of which we
reproduce the sum over instantons.

We now recall the ghost number conservation laws. The (virtual)
dimension of the moduli space of maps of degree $d$ from
the genus $g$ Riemann surface with
$s$ punctures onto $CP^1$ is given by
$$
{\rm dim}\, {\cal M}_{g,s}(CP^1;d)=2d+2(g-1)+s.
\eqno\eq
$$
The factor 2 in front of $d$ stands for
$c_1(CP^1)=2$. When all couplings vanish, genus $g$ correlation functions
$$
\bra \prod_{i=1}^s \S_{n_i}(\phi_{\alpha_i})\ket_g
\eqno\eq
$$
receive contributions from instantons which satisfy the conservation law
$$
\sum_{i=1}^s (n_i+q_{\alpha_i})=2d+2(g-1)+s,
\eqno\eq
$$
where $q_\alpha$ is the $U(1)$ charge of the field $\phi_\alpha$. Thus at
each genus $g$ contributions come from instantons of a definite degree
unlike the Calabi-Yau manifold case where the dimension of the moduli space is
independent of the degree of instantons and instantons of all possible
degrees contribute to a correlation function.

\chapter{Intersection numbers}

Let us first derive Ward identities of our model \partition\
by varying the matrix eigenvalues as
$\lambda_i \rightarrow \lambda_i+\varepsilon \lambda_i^{m+1}$. We find
$$
L_{-1}Z=\Big( -\pder{0,P}+\sum_{n=1}nt_{n,\alpha}\pder{n-1,\alpha}
+N^2\C \T \Big) Z=0,
\eqn\elmin
$$
$$
\eqalign{
L_0Z=\Big( \sum_{n=1} & nt_{n-1,Q}\pder{n-1,Q} -\pder{1,P}
+\sum_{n=1}nt_{n,P}\pder{n,P}  \cr
&-2\pder{0,Q}+2\sum_{n=1}nt_{n,P}\pder{n-1,Q}
+N^2\C^2\Big) Z=0,  \cr}
\eqn\elzero
$$
$$
\eqalign{
L_mZ = & \Big( -\pder{m+1,P}+\sum_{n=1}nt_{n,P}\pder{n+m,P}
+\sum_{n=1}(n+m)t_{n-1,Q}\pder{n+m-1,Q}  \cr
&-2(m+1)c_{m+1}\pder{m,Q}
+2\sum_{n=0}(n+m)(1+n(c_{n+m}-c_n))t_{n,P}\pder{n+m-1,Q}   \cr
&+{1 \over N^2}\sum_{r=1}^{m-1}r(m-r)
{\partial^2 \over \partial t_{m-r-1,Q}\partial t_{r-1,Q}} \Big)Z=0,
\hskip5mm m=1,2,\cdots . \cr}
\eqn\elm
$$
The operators $\{ L_m \}$ form a Virasoro algebra
$$
[L_m,L_n]=(m-n)L_{m+n}, \hskip5mm m,n \geq -1.
\eqno\eq
$$
The $L_0$ equation for the $CP^1$ model has also been derived by Hori
using the method of intersection theory without referring to the matrix
model
\REF\H{K. Hori, ``Constraints for topological strings in $D\geq 1$'',
UT-694, hep-th/9411135, to appear in Nucl. Phys. B.}
[\H].
The special feature of the above
operators is the presence of the mixing terms of the
form $t_{n,P}\partial /\partial t_{n+m-1,Q}$ which arise due to the
lack of scale invariance of the action \action.
The factor 2 in front of these terms is
identified as the 1st Chern class $c_1(CP^1)=2$.

If one switches off the coupling constants except $\C$, $\T$ and $t_{1,P}$,
one finds the genus $g=0$ free energy
$$
F_0=\half {\C^2\T \over 1-t_{1,P}}+(1-t_{1,P})^2e^{\T\over 1-t_{1,P}}
\eqno\eq
$$
by solving $L_{-1}$, $L_0$ equations and recovers \zerofree\ in the
small phase space.

Let us next consider other Virasoro operators and check them in a simple
case where we put all couplings
to zero except $\C$. We find, for instance,
$$
L_1Z=0 \Longrightarrow
-\bra \S_2(P)\ket_0-6\bra \S_1(Q)\ket_0+2\C \bra Q\ket_0 =0,
\eqn\elone
$$
$$
L_2Z=0 \Longrightarrow -\bra \S_3(P)\ket_0-11\bra \S_2(Q)\ket_0 +
4\C \bra \S_1(Q)\ket_0+\bra Q\ket_0 \bra Q\ket_0=0,
\eqn\eltwo
$$
$$
L_3Z=0 \Longrightarrow -\bra \S_4(P)\ket_0-{50 \over 3}\bra \S_3(Q)\ket_0
+6 \C \bra \S_2(Q)\ket_0 +4\bra \S_1(Q)\ket_0 \bra Q\ket_0 =0,
\eqn\elthree
$$
where $\bra \cdots \ket_0$ denotes the $g=0$ expectation value. We may
simply set $\C =0$ or take a suitable number of derivatives in $\C$ and
set $\C =0$ in the above equations.
${\partial \over \partial \C}|_{\C =0}$ of \elone\ gives, for instance,
$$
\eqalign{
-\bra \S_2(P)P\ket_0-6\bra \S_1(Q)P\ket_0 +2\bra Q\ket_0
&=-2\bra \S_1(P)\ket_0-6\bra Q\ket_0 +2\bra Q\ket_0  \cr
&=-2\cdot (-2)-4=0,  \cr}
\eqno\eq
$$
where we have used $\bra Q^n\ket_0 =1$ for $n=0,1,\cdots$ and
the puncture equation
$$
\bra P\prod_{i=1}^s \S_{n_i}(\phi_{\alpha_i}) \ket_0
=\sum_i n_i\bra \S_{n_1}(\phi_{\alpha_1})\cdots
\S_{n_{i-1}}(\phi_{\alpha_i}) \cdots
\S_{n_s}(\phi_{\alpha_s})\ket_0.
\eqno\eq
$$
The dilaton equation
$$
\bra \S_1(P)\prod_{i=1}^s \S_{n_i}(\phi_{\alpha_i}) \ket_g
=(2g-2+s)\bra \prod_{i=1}^s \S_{n_i}(\phi_{\alpha_i}) \ket_g
\eqno\eq
$$
has also been used (at $g=0$).
Similarly
${\partial^2 \over \partial \C^2}|_{\C =0}$ of \eltwo\ gives
$$
-\bra \S_3(P)PP\ket_0 -11\bra \S_2(Q)PP\ket_0+8\bra \S_1(Q)P\ket_0
+2\bra QPP\ket_0 \bra Q\ket_0
=-3 \cdot 2\cdot (-2)-11\cdot 2+8+2=0.
\eqno\eq
$$
We may also check
$$
\eqalign{
&{\partial^3 \over \partial \C^3}\Big|_{\C =0} \  \hbox{of \elthree\ }  \cr
=& -\bra \S_4(P)PPP\ket_0 -{50 \over 3}\bra \S_3(Q)PPP \ket_0
+6\cdot 3 \bra \S_2(Q)PP\ket_0  \cr
& \hskip15mm +3\cdot 4 \bra \S_1(Q)P \ket_0 \bra PPQ \ket_0
+4\bra \S_1(Q)PPP\ket_0 \bra Q \ket_0  \cr
=& -4\cdot 3\cdot 2\cdot (-2)-{50 \over 3}\cdot 3\cdot 2\cdot 1
+6\cdot 3\cdot 2 \cdot 1 +12 \cdot 1+4\cdot 1  \cr
=& 0.  \cr}
\eqno\eq
$$

If one wants to verify \elone-\elthree\ directly without taking derivatives
in $\C$, one has to prepare more geometrical data on correlation
functions. At $g=0$ intersection numbers are calculated by means of a
set of recursion relations which are derived by combining topological
recursion relation \toprec\ and Hori's equation for the intersection of
the K\"ahler class. Hori's relation [\H] reads as
$$
\bra Q \prod_{i=1}^s \S_{n_i}(\phi_{\alpha_i}) \ket_{0,d}
=d \bra \prod_{i=1}^s \S_{n_i}(\phi_{\alpha_i}) \ket_{0,d}
+\sum_{i=1}^s  n_i\bra \S_{n_1}(\phi_{\alpha_1})
\cdots \S_{n_{i-1}}(\phi_{{\alpha_i}+1}) \cdots
\S_{n_s}(\phi_{\alpha_s})\ket_{0,d} ,
\eqn\hori
$$
where $\bra \cdots \ket_{0,d}$ denotes the contribution of the degree-$d$
instanton to genus zero correlation functions.
By integrating \hori\ one finds
$$
\bra Q\ket_{0,d}=d\bra 1\ket_{0,d}+\sum_m mt_{m,\alpha}\pder{m-1,\alpha+1}
\bra 1\ket_{0,d} ,
\eqno\eq
$$
or
$$
\sum_m m\tilde t_{m,\alpha}\pder{m-1,\alpha+1} \bra 1\ket_{0,d}
=-d \bra 1\ket_{0,d} ,
\eqn\horider
$$
where we have defined
$\tilde t_{m,\alpha}=t_{m,\alpha}-\delta_{m,1}\delta_{\alpha,P}$.
Then we have
$$
\eqalign{
\Big( \sum_m m & \tilde t_{m,\alpha}\pder{m-1,\alpha+1}\Big)^2 \bra 1\ket_{0,d}
=d^2 \bra 1\ket_{0,d}    \cr
=& \sum m_1\tilde t_{m_1,\gamma}m_2 \tilde t_{m_2,\delta}
\bra \S_{m_1-1}(\phi_{\gamma +1})\S_{m_2-1}(\phi_{\delta +1})\ket_{0,d}  \cr
&+\sum m(m+1)t_{m,\gamma}\bra \S_{m-2}(\phi_{\gamma +2})\ket_{0,d} . \cr}
\eqn\horidouble
$$
The last term in the above equation can be dropped
in the $CP^1$ case (only two primaries exist). By taking the derivative
in $t_{n,\alpha}$ of \horidouble\
$$
\eqalign{
& d^2\bra \S_n(\phi_\alpha)\ket_{0,d}   \cr
=& \sum m_1\tilde t_{m_1,\gamma}m_2 \tilde t_{m_2,\delta}
\bra \S_n(\phi_\alpha) \S_{m_1-1}(\phi_{\gamma +1})
\S_{m_2-1}(\phi_{\delta +1})\ket_{0,d}   \cr
&+2n \sum m \tilde t_{m,\gamma}
\bra \S_{n-1}(\phi_{\alpha+1})\S_{m-1}(\phi_{\gamma +1})\ket_{0,d}  \cr
=& \sum \sum_{d_1+d_2=d}
m_1\tilde t_{m_1,\gamma}m_2 \tilde t_{m_2,\delta}
n\bra \S_{n-1}(\phi_\alpha)\phi_\beta \ket_{0,d_1}
\bra \phi^\beta \S_{m_1-1}(\phi_{\gamma +1})
\S_{m_2-1}(\phi_{\delta +1})\ket_{0,d_2}   \cr
&+2n \sum m \tilde t_{m,\gamma}
\bra \S_{n-1}(\phi_{\alpha+1})\S_{m-1}(\phi_{\gamma +1})\ket_{0,d}  \cr
=&\sum_{d_1+d_2=d}(d_2)^2n \bra \S_{n-1}(\phi_\alpha)\phi_\beta \ket_{0,d_1}
\bra \phi^\beta \ket_{0,d_2}-2nd \bra\S_{n-1}(\phi_{\alpha+1})\ket_{0,d} . \cr}
\eqno\eq
$$
Going from the 1st to 2nd line
we have used the topological recursion relation and used \horider\
in going from the 2nd to 3rd line.
Thus we have obtained
a basic recursion relation
$$
d^2 \bra \S_n(\phi_\alpha)\ket_{0,d}
=-2nd\bra \S_{n-1}(\phi_{\alpha+1})\ket_{0,d}
+\sum_{d_1+d_2=d}n(d_2)^2 \bra \S_{n-1}(\phi_\alpha)\phi_\beta \ket_{0,d_1}
\bra \phi^\beta \ket_{0,d_2}
\eqn\toprechori
$$
valid in the large phase space. \toprechori\ may be derived in a more
intrinsic manner using the method of
algebraic geometry. Derivation is given in Appendix B.

If one puts all couplings to zero and recalls
$\bra P\ket_{0,d}=\bra Q \ket_{0,d}=0$
except $\bra Q\ket_{0,1}=1$, one finds
$$
\eqalign{
(m+1)^2 \bra \S_{2m}(Q) \ket_{0,m+1}=& 2m \bra \S_{2m-1}(Q)P\ket_{0,m}
=2m(2m-1)\bra \S_{2m-2}(Q)\ket_{0,m} ,   \cr
(m+1)^2 \bra \S_{2m+1}(P) \ket_{0,m+1}
=&-2(2m+1)(m+1)\bra \S_{2m}(Q)\ket_{0,m+1}  \cr
&+(2m+1)2m\bra \S_{2m-1}(P)\ket_{0,m} .   \cr}
\eqno\eq
$$
(Note that due to the ghost number conservation
$\bra \S_{2m}(Q)\ket_{0,d}$ and $\bra \S_{2m+1}(P)\ket_{0,d}$ are
non-vanishing only for $d=m+1$).
These equations are easily solved and yield the data
$$
\eqalign{
\bra \S_{2m}(Q)\ket_{0,m+1}&={(2m)! \over (m+1)!(m+1)!},  \cr
\bra \S_{2m+1}(P)\ket_{0,m+1}&=-2c_{m+1}{(2m+1)! \over (m+1)!(m+1)!} .\cr}
\eqn\intnumber
$$

Now one may check \eltwo\ at $\C =0$,
$$
-\Big( -{9 \over 2}\Big)-11\cdot \half+1\cdot 1=0.
\eqno\eq
$$
One may also verify $\partial /\partial \C$ of \elthree\ at $\C =0$,
$$
-4\cdot \Big( -{9 \over 2}\Big)-{50 \over 3}\cdot 3\cdot \half
+6 \cdot \half+4\cdot 1\cdot 1=0.
\eqno\eq
$$

We can also discuss our model at genus $g=1$ and compare its predictions
with geometrical data.
If we put all couplings to zero except $\C$, we find
$$
L_{-1}Z=0 \Longrightarrow \bra P\ket_1=0,
\eqno\eq
$$
$$
L_{0}Z=0 \Longrightarrow \bra \S_1(P)\ket_1+2\bra Q\ket_1=0.
\eqn\torusonept
$$
Here $\bra \cdots \ket_1$ denotes the genus $g=1$ expectation value.
\torusonept\ is consistent with the geometrical data [\H]
$$
\eqalign{
\bra \S_1(P)\ket_1 &={1 \over 24} \chi(CP^1)={1\over 12},  \cr
\bra Q\ket_1 &=-{1 \over 24}\int_{CP^1} \omega
=-{1 \over 24}. \cr}
\eqn\data
$$
Other Virasoro operators predict
$$
L_1Z=0 \Longrightarrow -\bra \S_2(P)\ket_1-6\bra \S_1(Q)\ket_1
+2\C \bra Q\ket_1=0,
\eqno\eq
$$
$$
L_2Z=0 \Longrightarrow -\bra \S_3(P)\ket_1-11\bra \S_2(Q)\ket_1+\bra QQ\ket_0
+2\bra Q\ket_1\bra Q\ket_0+4\C \bra \S_1(Q)\ket_1=0,
\eqno\eq
$$
$$
\eqalign{
L_3Z=0 \Longrightarrow &-\bra \S_4(P)\ket_1-{50 \over 3}\bra \S_3(Q)\ket_1
+4 \bra \S_1(Q)Q\ket_0   \cr
&+4\bra Q\ket_0\bra \S_1(Q)\ket_1
+4\bra Q\ket_1\bra \S_1(Q)\ket_0+6 \C \bra \S_2(Q)\ket_1, \cr}
\eqno\eq
$$
etc.. It is easy to
see that these relations are all satisfied as in the $g=0$ case.

There is in fact a subtlety associated with the $g=1$ free energy. From
the definition of the partition function \partition\ and
the fact that the size
of the matrix $M$ being equal to $(N\times t_{0,P})^2$, it is possible to
derive a relation
$$
\Big(D-N{\partial \over \partial N}\Big)F = 0,
\eqn\euler
$$
where $F=\log Z$ and $D=\sum_{n=0} \tilde t_{n,\alpha}
{\partial \over \partial t_{n,\alpha}}$.
If the free energy has an expansion without a term proportional to $\log N$,
\euler\ would imply $\bra \sigma(P) \ket_1 = 0$ at zero couplings in
contradiction to the data \data. Thus $F$ must contain
a $\log N$ term as
$$
F=\sum_{g=0} N^{2-2g}F_g -{1 \over 12}\log N
\eqn\mfr
$$
in the $1/N$ expansion. \mfr\ reproduces \data.

The presence of the $\log$ term is related to the issue of whether the
genus-$1$ free energy has an expression suggestive of a string field theory
$$
F_1 = {1 \over 24} \log \det u_{\alpha \beta},
\eqn\onefr
$$
where $u_{\alpha \beta}=\partial^3 F_0/\partial t_{0,P}\partial t_{0,\alpha}
\partial t_{0,\beta}$. Unlike the case of minimal models where \onefr\ holds
\REF\IZ{C. Itzykson and J.-B. Zuber, \ijmp{7} (1992) 5661.}
\REF\EYY{T. Eguchi, Y. Yamada and S.-K. Yang, ``On the genus expansion
in the topological string theory'', UTHEP-275, hep-th/9405106,
to appear in Rev. Math. Phys.}
[\DW,\IZ,\EYY],
it may not be valid in the case of $\sigma$-models
\REF\Wsur{E. Witten, Surveys in Diff. Geom. 1 (1991) 243.}
[\Wsur].
In the $CP^1$ model we have instead a relation
$$
F_1 = {1 \over 24} \log \det u_{\alpha \beta}-{1 \over 24} u,
\hskip5mm (\alpha,\beta = 1,2)
\eqn\monefr
$$
where the extra term reproduces $\bra Q \ket_1 = -{1 \over 24}$.
In our previous article [\EY] we wrote down flow equations which did
not contain the 2nd term in the RHS of \monefr. We present corrected flow
equations in the Appendix A.

\chapter{Landau-Ginzburg formulation}

In the case of topological minimal models coupled to topological gravity
at genus $g=0$
Landau-Ginzburg (LG) description of the system has been developed where
the scalar Lax operator of the dispersionless KP hierarchy
plays the role of the superpotential
\REF\DVV{R. Dijkgraaf, E. Verlinde and H. Verlinde, \np{352} (1991) 59.}
[\DVV].
In the following we shall show that it is also possible to develop a LG
formulation in the case of the topological $CP^1$ model using the Lax
operator of the Toda hierarchy as the superpotential.
Throughout this section we restrict ourselves to the small phase space, i.e.
$u=\T$ and $v=\C$. The Lax operator is then given by
$$
L=p+\C+e^{\T} p^{-1}.
\eqn\superpot
$$
If one regards $L$ as the superpotential, its extrema give the vacuum states
$$
\partial_p L=0 \  \Longrightarrow \  p^2=e^{\T} .
\eqno\eq
$$
If the K\"ahler class $Q$ is described by $p$, this gives the relation
$Q \cdot Q=e^{\T}$ of
the quantum cohomology ring of $CP^1$. We thus make an identification
$$
P=1, \hskip7mm  Q=p.
\eqn\field
$$
Note that the variable $p$ itself is not the Landau-Ginzburg field
since the Poisson bracket of the Toda theory \poissonbra\ has an
unusual form with
an extra factor of $p$ in the RHS. We may instead regard $X=\log p$ as the
LG variable. The superpotential \superpot\ is then rewritten as
$$
L=\exp X + t_{0,P} + e^{t_{0,Q}}\exp(-X).
\eqno\eq
$$
It is of the form of the sine-Gordon potential.
Because of the change of variable $p \rightarrow X$ there appear
Jacobian factors which one must take into account.
The residue formula is now given by
$$
\bra ABC \ket =\oint dp {A(p)B(p)C(p) \over p^2 \partial_p L},
\eqn\residue
$$
where the integration contour is taken around $p=\infty$.
Note that the denominator equals
$$
p^2\partial_p L = p^2 - e^{t_{0,Q}}
\eqno\eq
$$
which looks as if we were dealing with the case of the $A_1$ minimal model.
It is easy to check the flatness of the metric
$$
\bra P \phi_\alpha \phi_\beta \ket =\eta_{\alpha\beta},
\eqno\eq
$$
where $\eta_{PQ}=\eta_{QP}=1, \ \eta_{PP}=\eta_{QQ}=0$.

Our next step is to describe gravitational descendants $\S_n(\phi_\alpha),\
\phi_\alpha=P,Q$.
In the case of the topological minimal models coupled to topological gravity
it is possible to adopt the matter picture and express
gravitational descendants using only the matter fields
\REF\EKYY{T. Eguchi, H. Kanno, Y. Yamada and S.-K. Yang,
\pl{305} (1993) 235.}
\REF\L{A. Losev, Theor. Math. Phys. {\bf 95} (1993) 595.}
[\EKYY,\L].
It turns out that also in the case of $CP^1$ model it is possible to
describe gravitational descendants using only the LG field.
We postulate
$$
\eqalign{
\S_n(P) &=2n [L^{n-1}(\log L-c_{n-1})p\partial_p L]_+ \ ,  \cr
\S_n(Q) &=[L^np\partial_p L]_+ \ , \cr}
\eqn\descen
$$
with $n=0,1,2,\cdots$. Note that $\S_n(Q)$ is a polynomial of order $n+1$
in $p$, while $\S_n(P)$ is an infinite series in the variable $p$.
We remark that $\S_0(Q)=p$ from \descen\
in agreement with \field.
On the other hand, $\S_0(P)$ differs from $1$, however, the difference is
a BRST exact term.

In order to check our LG formulation we look at the flow
equations of the $CP^1$ model at genus zero.
Let us recall the flow equations of Section 2,
$$
\eqalign{
{\partial L \over \partial t_{n,P}} &=
2\{ [L^{n}(\log L-c_{n})]_+, L \} , \cr
{\partial L \over \partial t_{n,Q}} &=
{1 \over n+1} \{[L^{n+1}]_+,L \}. \cr}
\eqn\flow
$$
Comparing the coefficients of $p^{-1}$ on both sides we find
$$
{\partial u \over \partial t_{n,\alpha}}=
{\partial  \over \partial t_{0,P}} \bra \S_n(\phi_\alpha)P \ket ,
\eqno\eq
$$
where
$$
\eqalign{
\bra \S_n(P)P \ket &=2 {\rm res} (L^{n}(\log L-c_{n})/p) , \cr
\bra \S_n(Q)P \ket &={1 \over n+1} {\rm res} (L^{n+1}/p) . \cr}
\eqn\twopt
$$
These are the analogues of the Gelfand-Dikii potentials of the KP
hierarchy. Here ``res'' means to take the coefficient of $p^{-1}$. We denote
$\bra \S_n(\phi_\alpha)P \ket =R_{n,\alpha}$ henceforth.
Integrating \twopt\ over $\C$ we find
$$
\eqalign{
\bra \S_n(P) \ket &= {1 \over n+1}R_{n+1,P}=
{2 \over n+1}{\rm res} (L^{n+1}(\log L - c_{n+1})/p), \cr
\bra \S_n(Q) \ket &={1 \over n+1}R_{n+1,Q}
={1 \over (n+2)(n+1)} {\rm res} (L^{n+2}/p). \cr}
\eqn\onept
$$
3-point functions $\bra \S_n(\phi_\alpha)P \phi_\beta \ket$ are
calculated either by taking derivatives of \twopt\ or by
using the residue formula \residue. We find the same results.

We now discuss how various types of recursion relations can be recovered
in the LG approach. First of all, notice that
$$
{\partial  \over \partial t_{0,P}} \S_n(\phi_\alpha)=n\S_{n-1} (\phi_\alpha),
\hskip7mm n \geq 1.
\eqno\eq
$$
This relation leads to the puncture equation
$$
\eqalign{
\langle P & \sigma_{n_1}(\phi_\alpha)
\sigma_{n_2}(\phi_\beta)\sigma_{n_3}(\phi_\gamma)\rangle
={\partial \over \partial \C} \langle  \sigma_{n_1}(\phi_\alpha)
\sigma_{n_2}(\phi_\beta)\sigma_{n_3}(\phi_\gamma)\rangle  \cr
&={\partial \over \partial \C} \oint dp {\sigma_{n_1}(\phi_\alpha(p))
\sigma_{n_2}(\phi_\beta(p))\sigma_{n_3}(\phi_\gamma(p))
\over p^2 \partial_p L} \cr
&=n_1 \langle \sigma_{n_1-1}(\phi_\alpha)
\sigma_{n_2}(\phi_\beta)\sigma_{n_3}(\phi_\gamma)\rangle+
n_2 \langle \sigma_{n_1}(\phi_\alpha)
\sigma_{n_2-1}(\phi_\beta)\sigma_{n_3}(\phi_\gamma)\rangle \cr
&~~~+n_3 \langle \sigma_{n_1}(\phi_\alpha)
\sigma_{n_2}(\phi_\beta)\sigma_{n_3-1}(\phi_\gamma)\rangle . \cr}
\eqno\eq
$$

In order to verify the topological recursion relation we first
make a decomposition of gravitational descendants into primary components
and BRST exact pieces
$$
\eqalign{
\S_n(P) &=2np^2 \partial_pL[L^{n-1}(\log L-c_{n-1})/p]_+
+\sum_{\alpha,\beta} \eta^{\alpha\beta}
{\partial R_{n,P} \over \partial t_{0,\alpha}} \phi_\beta \ , \cr
\S_n(Q) &=p^2 \partial_pL [L^{n}/p]_+
+\sum_{\alpha,\beta} \eta^{\alpha\beta}
{\partial R_{n,Q} \over \partial t_{0,\alpha}} \phi_\beta \ . \cr}
\eqn\decomp
$$
Derivation of \decomp\ is in parallel with that explained in [\EKYY].
Thus
$$
\eqalign{
&\langle \sigma_n(\phi_\alpha)X Y \rangle
=\oint dp {\sigma_n(\phi_\alpha)X Y \over p^2 \partial_p L}
=\sum_{\beta ,\gamma} {\partial \over \partial t_{0,\beta}} R_{n,\alpha}
\eta^{\beta\gamma} \langle \phi_\gamma XY \rangle  \cr
& =\sum_{\beta ,\gamma} n {\partial \over \partial t_{0,\beta}}
 \langle \sigma_{n-1}(\phi_\alpha)\rangle \eta^{\beta\gamma}
\langle\phi_\gamma XY \rangle
=\sum_{\beta ,\gamma} n \langle \sigma_{n-1}(\phi_\alpha)\phi_{\beta}\rangle
\eta^{\beta\gamma} \langle\phi_{\gamma}XY \rangle \ , \cr}
\eqn\toprec
$$
where we have used \onept. This is the desired relation.

Now, turn off $\C$ and make a change of variable $p=e^{\T /2}{\bar p}$, then
$$
\S_n(\phi_\alpha)=e^{(n+q_\alpha)\T /2}{\bar \sigma}_n(\phi_\alpha),
\eqn\reddescen
$$
where the $U(1)$ charges are $q_P=0,\ q_Q=1$ and
$$
\eqalign{
{\bar \sigma}_n(P)
&=2n [{\bar L}^{(n-1)}(\log L-c_{n-1}){\bar p}\partial_{{\bar p}} {\bar L}]_
+ \ ,\cr
{\bar \sigma}_n(Q) &=[{\bar L}^n {\bar p}\partial_{{\bar p}} {\bar L}]_+ \ ,
\cr}
\eqno\eq
$$
with
$$
\eqalign{
{\bar L} &={\bar p}+{\bar p}^{-1}, \cr
\log L &=\half \T+\half \log (1+{\bar p}^2)+\half \log(1+{\bar p}^{-2}). \cr}
\eqno\eq
$$
Notice that
$$
{\partial \over \partial\T} {\bar \sigma}_n(P)=n {\bar \sigma}_{n-1}(Q),
\hskip1cm
{\partial \over \partial\T} {\bar \sigma}_n(Q)=0.
\eqn\qderiv
$$
The degree-$d$ instanton contribution to the 3-point function
$\bra \S_{n_1}(\phi_{\alpha_1}) \S_{n_2}(\phi_{\alpha_2})
\S_{n_3}(\phi_{\alpha_3}) \ket_d$
is given by
$$
\bra \S_{n_1}(\phi_{\alpha_1}) \S_{n_2}(\phi_{\alpha_2})
\S_{n_3}(\phi_{\alpha_3}) \ket_d
=e^{d\T} \oint d{\bar p}
{{\bar \sigma}_{n_1}(\phi_{\alpha_1}) {\bar \sigma}_{n_2}(\phi_{\alpha_2})
{\bar \sigma}_{n_3}(\phi_{\alpha_3}) \over {\bar p}^2 \partial_{{\bar p}}
{\bar L}},
\eqn\three
$$
where we have used the ghost number conservation
$\sum_{i=1}^3 (n_i+q_{\alpha_i}-1)=2d-2$.
Combining \three\ and \qderiv\ we find
$$
\eqalign{
\bra Q & \S_{n_1}(\phi_{\alpha_1}) \S_{n_2}(\phi_{\alpha_2})
\S_{n_3}(\phi_{\alpha_3}) \ket_d
={\partial \over \partial\T}
\bra \S_{n_1}(\phi_{\alpha_1}) \S_{n_2}(\phi_{\alpha_2})
\S_{n_3}(\phi_{\alpha_3}) \ket_d   \cr
&=d e^{d\T} \oint {d{\bar p} \over {\bar p}^2 \partial_{{\bar p}}{\bar L}}
\prod_{i=1}^3 {\bar \sigma}_{n_i}(\phi_{\alpha_i})
+e^{d\T}\oint {d{\bar p}\over {\bar p}^2 \partial_{{\bar p}}{\bar L}}
{\partial \over \partial\T}
\prod_{i=1}^3 {\bar \sigma}_{n_i}(\phi_{\alpha_i}) \cr
&=d \bra \S_{n_1}(\phi_{\alpha_1}) \S_{n_2}(\phi_{\alpha_2})
\S_{n_3}(\phi_{\alpha_3}) \ket_d+
n_1 \bra \S_{n_1-1}(Q) \S_{n_2}(\phi_{\alpha_2})
\S_{n_3}(\phi_{\alpha_3}) \ket_d  \delta_{\alpha_1P} \cr
& \ \ +n_2 \bra \S_{n_1}(\phi_{\alpha_1}) \S_{n_2-1}(Q)
\S_{n_3}(\phi_{\alpha_3}) \ket_d \delta_{\alpha_2P}
+n_3 \bra \S_{n_1}(\phi_{\alpha_1}) \S_{n_2}(\phi_{\alpha_2})
\S_{n_3-1}(Q) \ket_d \delta_{\alpha_3P}.  \cr}
\eqno\eq
$$
This is the relation \hori\ for the insertion of the K\"ahler class.

Finally we present some sample calculations.
First we compute $\bra \S_3(P)\S_1(P)\S_1(P) \ket$ at
$\C=0,\ \T \not= 0$. We write
$$
\S_3(P)=A(p)+B(p), \hskip10mm  \S_1(P)\S_1(P)=C(p)+D(p),
\eqno\eq
$$
where
$$
\eqalign{
A(p)&=\half (e^{-\T}p^2+\T-c_2)(p^3+e^{\T})p,
\hskip10mm B(p)=\sum_{\ell=1}^\infty B_\ell ~ p^\ell,  \cr
C(p)&=\T^2p^2, \hskip10mm
D(p)=\sum_{\ell=1}^\infty D_\ell ~ p^\ell   \cr}
\eqno\eq
$$
and the coefficients $B_\ell, D_\ell$ are easily read off from \descen.
Expanding $1/p^2\partial_pL$ around $p=\infty$ we have
$$
\eqalign{
\bra \S_3(P) &\S_1(P)\S_1(P) \ket  \cr
& =\oint dp \sum_{k=0}^\infty {e^{k\T} \over p^{2k+2}}
[A(p)C(p)+A(p)D(p)+B(p)C(p)+B(p)D(p)]. \cr}
\eqno\eq
$$
Explicit calculations show that the contributions from the last three terms
cancel out among themselves. The first term then yields
$$
\bra \S_3(P)\S_1(P)\S_1(P) \ket=6 \T^2 (\T-2)e^{2\T}.
\eqno\eq
$$
This result can be reproduced by a computation using \flow\ and \toprec.

Setting $\C=\T=0$ we next examine 1-point functions. Using the binomial
expansion of powers of $L=p+p^{-1}$ in \onept\ we immediately recover the
previous result \intnumber\
$$
\bra \S_{2m}(Q) \ket ={(2m)! \over (m+1)!(m+1)!}.
\eqno\eq
$$
Similarly $\bra \S_{2m+1}(P) \ket$ can be evaluated as
$$
\eqalign{
\bra \S_{2m+1}(P) \ket
&={1 \over m+1} \oint {dp \over p}L^{2m+2}(\log L-c_{2m+2}) \cr
&={1 \over m+1} \oint {dp \over p}L^{2m+2}(\log(1+p^{-2})-c_{2m+2})  \cr
&={1 \over m+1} \oint {dx \over (x-1)^{2m+2}}(\log x-c_{2m+2}) \cr
&={2 (2m+1)! \over (m+1)!(m+1)!} \oint dx
{x^{m+1} \over x-1}(\log x-c_{m+1}),  \cr}
\eqno\eq
$$
where we have made a change of variable $1+p^{-2}=x$ and repeated partial
integrations. Then by picking up a pole
at $x=1$ we reproduce \intnumber\
$$
\bra \S_{2m+1}(P) \ket =-2c_{m+1}{(2m+1)! \over (m+1)!(m+1)!}.
\eqno\eq
$$

As we have noted earlier, the superpotential of the $CP^1$ model
has the form of the
sine-Gordon theory when expressed in terms of the LG field $X$. This seems to
indicate the equivalence of the $CP^1$ and the $N=2$ supersymmetric sine-Gordon
theories at least in their topological versions.
The correspondence between these
theories have been known for some time based on the agreement of the particle
spectra and scattering matrices [\CV,\FI].
Our results seem to provide further evidence
for their equivalence. It will be interesting to see if we can make the
correspondence more precise and find out, in particular,
how a super-renormalizable
sine-Gordon theory may be turned into an asymptotically free theory possibly
at some special value of the coupling constant.

The relation between our coupling constant $\T$
and those of the physical theory $g^2/4\pi$, $\theta$ is given by
$$
\T=-\Big({4\pi \over g^2} + i \theta\Big) .
\eqno\eq
$$
\REF\MPS{A.Y. Morozov, A.M. Perelomov and M.A. Shifman, \np{248} (1984) 279.}
In asymptotically free theories the coupling constant is
replaced by the $\Lambda$-parameter via the dimensional transmutation
$e^{-4\pi /g^2}=\Lambda^b$ with
$b$ being the one-loop $\beta$-function. For the $N=2$ $CP^n$ $\S$-model
$b$ is equal to the 1st Chern class, $b=n+1$ [\MPS]. Thus we have (putting
$\theta=0$)
$$
e^{\T}=\Lambda^2
\eqno\eq
$$
and the superpotential reads after shifting $X \rightarrow X+\T /2$
$$
L=\Lambda  (e^X+ e^{-X}).
\eqno\eq
$$
Clearly $\Lambda$ corresponds to the sine-Gordon mass parameter.

\chapter{Generalizations and Discussions}

In this paper we have discussed the topological $CP^1$ model in some details.
We have described its integrable structure, matrix model realization and
Landau-Ginzburg formulation. Predictions of the matrix model on the
intersection numbers (Gromov-Witten invariants) on the moduli space
$\overline{\cal M}_{g,s}(CP^1;d)$ agree with those obtained by using the method
of algebraic geometry and also by residue integrals of the LG superpotential
at $g=0,\ 1$. Thus the model in fact seems consistent and reproduces the
sum over instantons from the Riemann surfaces to $CP^1$. It will be important
to verify the predictions of the model at higher genera.

It is somewhat unexpected that the Virasoro type constraints also hold in the
$CP^1$ theory although the Virasoro operators have a peculiar form reflecting
the lack of scale invariance in the system. Unlike the case of minimal
models, however, we can not construct negative-index operators
$\{ L_{-n}, n=2,3,\cdots\}$ to supplement the original ones
$\{ L_m, m \geq -1\}$ into a full-fledged Virasoro algebra. In the $CP^1$
case only ``half'' of the algebra exists. A related issue is a possible
free boson/fermion description of the $CP^1$ system. While the form of
the Virasoro operators suggests free bosons associated with each set of
variables $\{ t_{n,P}\}$, $\{ t_{n,Q}\}$, it differs from that of the
canonical stress tensor and does not seem to have an obvious interpretation.
It will be an interesting problem to provide a space-time description of the
$CP^1$ model.

Another important problem is to extend our construction to $CP^n$ or other
Fano varieties. It turns out that in the case of $CP^2$ and $CP^3$ it is also
possible to develop a LG formulation; in the $CP^2$ case, for instance,
we assume
that the primary fields $P,Q,R$ are represented as
$$
P=1, \ \ \ \ Q=p+a, \ \ \ \  R= p^2+bp+c
\eqn\cptwoprim
$$
and satisfy the algebra
$$
\phi_i\phi_j=\bra \phi_i\phi_j\phi^k\ket \phi_k \ \ \ \ {\rm modulo} \ W'.
\eqn\cptwoope
$$
It is also assumed that the potential $W$ has a form
$$W=p^3+up^2+vp+w.
\eqn\cptwopot
$$
It is then possible to find a solution to the above set of equations
(by shifting $p$ we may generate other solutions)
$$
\eqalign{
&a=\bra QQQ\ket, \ \ \ b=\bra QQQ\ket, \ \ \ c=-\bra QQR\ket, \cr
&u=2\bra QQQ\ket, \ \ \ v=\bra QQQ\ket^2-2\bra QQR\ket,
\ \ \ w=-2\bra QQQ\ket\bra QQR\ket - \bra QRR\ket \cr}
\eqno\eq
$$
and we further predict
$$
\bra RRR\ket= -\bra QQQ\ket\bra QRR\ket + \bra QQR\ket^2.
\eqn\cptwoass
$$
\cptwoass\ is the equation which comes from the associativity of
the operator-product-expansion [\KM]. Note that when we switch off $t_P,t_R$,
$\bra QQQ\ket,\bra QQR\ket,\bra RRR\ket$ vanishes and $\bra QRR\ket=e^{t_Q}$
and the
above construction reduces to the familiar relations in quantum cohomology
\REF\V{C. Vafa, ``Topological mirrors and quantum rings'', in {\it Essays on
Mirror Manifolds}, S.-T.Yau ed., International Press Co., Hong Kong, 1992.}
\REF\I{K. Intriligator, \mpl{6} (1991) 3543.}
[\V,\I],
$$
\eqalign{
&P=1, \ \ \ Q=p, \ \ \ R=p^2 \cr
&W=p^3-e^{t_Q}. \cr}
\eqno\eq
$$

A similar construction works also for the $CP^3$ model and we deform
quantum cohomology relations. However, we do not
yet know the integrability structure behind these constructions.
(Recently Kanno-Ohta
\REF\KO{H. Kanno and Y. Ohta, ``Topological strings with scaling violation
and Toda lattice hierarchy'', hep-th/9502029.}
[\KO]
have introduced an interesting model whose superpotential
has a form $ap^2+\cdots +bp^{-1}$. The model, however, does not
describe $CP^2$ but a modified $CP^1$ theory coupled with some
minimal matter).

\vskip10mm
We would like to thank Y. Yamada for pointing out the structure of the
$g=1$ free energy in the $CP^1$ model. We also thank a comment by C. Itzykson
on the related issue. S.K.Y. has benefited from discussions with P. Fendley,
K. Intriligator and S.-J. Rey. Research of T.E. and S.K.Y. is supported
in part by Grant-in-Aid for Scientific Research on Priority Area 231
``Infinite Analysis'', Japan Ministry of Education.

\endpage

\Appendix{A}

Flow equations at higher genera of [\EY] do not reproduce the $g=1$ free
energy \monefr\
(the relation between the 2-point function $u=\bra PP\ket$ and the
free energy was not chosen properly at higher genera). We present in the
following the revised version of the flow equations.
$$
{\partial u \over \partial t_{0,Q}} = \bra PQ \ket ' = [v]'
\eqno\eq
$$
$$
\eqalign{
{\partial v \over \partial t_{0,Q}} &= \bra QQ \ket ' \cr
&= \Big[e^u + {1 \over N^2} {1 \over 12}u''e^u
+{1 \over N^4}
\Big( {1 \over 360}u^{(4)} +{1 \over 288 }u''^2 \Big) e^u +\cdots \Big]' \cr}
\eqno\eq
$$
$$
\eqalign{
{\partial u \over \partial t_{1,P}} &= \bra \Desc{1}{P} P \ket ' \cr
&= \Big[ uv+ {1 \over N^2} \Big( {1 \over 6} v''-{1 \over 12}v'u' \Big)  \cr
& \hskip10mm + {1 \over N^4}
\Big( -{1 \over 360} v^{(4)}+{1 \over 720}v'''u'+{1 \over 180}v''u''
+{1 \over 720} v'u''' \Big) +\cdots   \Big]' \cr}
\eqno\eq
$$
$$
\eqalign{
{\partial v \over \partial t_{1,P}} &= \bra \Desc{1}{P} Q \ket ' \cr
&=\Big[\half v^2+(u-1)e^u+
{1 \over N^2}
\Big( \Big( {1 \over 12} uu''+{1 \over 6}u''+{1 \over 12} u'^2 \Big)e^u
-{1 \over 24}v'^2   \Big)      \cr
& \hskip10mm + {1 \over N^4}
\Big( \Big( {1 \over 360}uu^{(4)}+{1 \over 90}u^{(4)}+{1 \over 120} u'u'''
-{1 \over 720} u'^2u''   \cr
& \hskip20mm
-{1 \over 720}u'^4+{1 \over 288}uu''^2+{19 \over 1440}u''^2 \Big)e^u
+{1 \over 720} v'''v'+{1 \over 360}v''^2 \Big) +\cdots \Big]'  \cr}
\eqno\eq
$$
$$
\eqalign{
{\partial u \over \partial t_{1,Q}} &= \bra \Desc{1}{Q} P \ket ' \cr
&= \Big[ \half v^2+e^u
+ {1 \over N^2} \Big( \Big( {1 \over 12} u'^2
+{1 \over 6} u''\Big)e^u-{1\over 24}v'^2 \Big)  \cr
&\hskip10mm
+ {1 \over N^4} \Big( \Big({1 \over 120} u^{(4)}+{1 \over 120} u'u'''
-{1 \over 720} u'^2u''-{1 \over 720} u'^4+{1\over 160} u''^2 \Big)e^u \cr
&\hskip20mm
+{1 \over 720} v'v'''+{1 \over 360} v''^2 \Big)  +\cdots \Big]' \cr}
\eqno\eq
$$
$$
\eqalign{
{\partial v \over \partial t_{1,Q}} &= \bra \Desc{1}{Q} Q \ket ' \cr
&= \Big[v e^u
+ {1 \over N^2} \Big( {1 \over 12} vu''+{1 \over 6}v'' \Big) e^u   \cr
& \hskip10mm + {1 \over N^4}
\Big({1 \over 360} vu^{(4)}+{1\over 288} vu''^2+{1 \over 72}v''u''
+{1 \over 120} v^{(4)} \Big)e^u  +\cdots \Big]'  \cr}
\eqno\eq
$$

\Appendix{B}

Any holomorphic map $f:\CP\to \CP$ of degree $d$ is expressed as
$$
f(z)=a {(z-b_1)\cdots(z-b_d) \over (z-c_1)\cdots(z-c_d)},
\eqno\eq
$$
where $b_i\ne c_j$ for every $i,j$, or is obtained as the limit of such an
expression as $b_i\to \infty, a\to 0$ (or $c_j\to \infty, a\to \infty$).
Two maps $f$ and $f'$ are equivalent when there is a fractional linear
transformation $g\in PSL(2,\bfC)$
$$
g(z)={\alpha z+\beta \over \gamma z+\delta}\qquad
\pmatrix{
\alpha & \beta\cr
\gamma & \delta\cr
}\in SL(2,\bfC),
\eqno\eq
$$
such that $f=f'\circ g$. This induces the action of $PSL(2,\bfC)$
on the parameters $b_i, c_j$:
$$
b_i\mapsto g(b_i),\qquad c_j\mapsto g(c_j)
\eqno\eq
$$
while $a$ is transformed in a complicated way.
A marked point $x$ is also transformed as
$$
x\mapsto g(x).
\eqno\eq
$$
Let $\Md$ be the moduli space of degree $d$ holomorphic maps $\CP\to\CP$
with one marked point on the world sheet. The expression
$$
s\,=\bigotimes_{1\leq i,j\leq d}\Big( \,{dx \over x-b_i}-{dx \over x-c_j}
\,\Big)
\eqno\eq
$$
is $PSL(2,\bfC)$-invariant and is symmetric under permutations
$b_i\mapsto b_{i'}, c_j\mapsto c_{j'}$.
Therefore, $s$ is a meromorphic section of the line bundle
${\cal L}^{\otimes d^2}$ over $\Md$ where ${\cal L}$ is the bundle of
cotangent space at the marked point.
It extends over the compactified moduli space $\bMd$
introduced in
\REF\K{M. Kontsevich, ``Enumeration of rational curves via torus actions'',
MPI preprint, hep-th/9405035.}
[\KM,\K].
One can figure out the first Chern class $c_1({\cal L}^{\otimes d^2})
=d^2c_1({\cal L})$ by looking at the locus of zeroes and poles of $s$.

\vskip0.3cm
\undertext{Poles}

The section $s$ has poles of order $d$ at the divisors $D_b$ and $D_c$
where $D_b$ (resp. $D_c$) is the locus of $x=b_i$ (resp. $x=c_i$)
for some $i$. In other words, $D_b$ (resp. $D_c$) is the locus
of configurations such that $x$ is mapped by $f$ to $0\in \CP$
(resp. $\infty\in \CP$). The Poincar\'e dual is thus
${[}D_b{]}={[}D_c{]}=\phi^*\omega$ where $\phi$ is the evaluation map
and $\omega$ is the K\"ahler form of volume $1$.

\vskip0.3cm
\undertext{Zeroes}

The section $s$ goes to zero as $b_i$ approaches to $c_j$.
Let us see what happens to the configuration when the $2k$ points
$b_1,\dots, b_k,c_1,\cdots, c_k$ converge into one point $x_*$:
$$
\left.
\eqalign{
b_i &=x_*+\beta_i\epsilon  \cr
c_i &=x_*+\gamma_i\epsilon  \cr}
\right\} \mathop{\longto}^{\epsilon\to 0}x_*\qquad i=1,\cdots,k.
\eqno\eq
$$
If we use $z$ as the coordinate on the world sheet, $f$ converges for
$z\ne x_*$ as
$$
f(z)\mathop{\longto}^{\epsilon\to 0}
a {(z-b_{k+1})\cdots(z-b_d) \over (z-c_{k+1})\cdots(z-c_d)}=: f_0(z).
\eqno\eq
$$
$f_0$ is generically a map of degree $d-k$.
If we use instead the coordinate $\zeta$ defined by
$$
z-x_*=\epsilon \zeta,
\eqno\eq
$$
$f(z)=\tilde{f}(\zeta)$ converges for $\zeta\ne\infty$ as
$$
\tilde{f}(\zeta)\mathop{\longto}^{\epsilon\to 0}
f_0(x_*)
{(\zeta-\beta_1)\cdots(\zeta -\beta_k) \over (\zeta-\gamma_1)\cdots
(\zeta -\gamma_k)}=:\tilde{f}_0(\zeta).
\eqno\eq
$$
$\tilde{f}_0$ is generically a map of degree $k$.
The limit can be identified as the configuration of genus zero
Riemann surface $\Sigma_0\cup\tilde{\Sigma}_0$
with a double point $\Sigma_0\cap \tilde{\Sigma}_0$
where $\Sigma_0$ is mapped to $\CP$ by a map of degree $d-k$
and $\tilde{\Sigma}_0$ is mapped to $\CP$ by a map of degree $k$.
(Generically,
$\Sigma_0$ is $\CP$ with coordinate $z$ that is mapped to $\CP$ by
$f_0$ and $\tilde{\Sigma}_0$ is $\CP$ with coordinate $\zeta$
that is mapped by $\tilde{f}_0$.
The double point $\Sigma_0\cap \tilde{\Sigma}_0$ ($z=x_*$ or $\zeta=\infty$)
is mapped to $f_0(x_*)$.)
We denote by $D_{d-k,k}$ the moduli space of such configurations
with a marked point in $\Sigma_0$ (the branch of degree $d-k$).
The moduli space $\bMd$ includes $D_{d-k,k}$ as a compactification divisor
to which the coordinate $\epsilon$ is transversal.
Since $s$ is proportional to $\epsilon^{k^2}$ as $\epsilon \to 0$,
it has zero of order $k^2$ at $D_{d-k,k}\subset\bMd$.

Thus, we see that
$$
d^2c_1({\cal L})=-2d\,\phi^*\omega+\sum_{k=1}^d k^2{[}D_{d-k,k}{]}.
\eqno\eq
$$
The same argument shows the similar relation
in the case with extra marked points. Let $S$ be the set of marks for them.
Denoting by $\sum_{X\cup Y=S}$ the sum over disjoint unions, we have
$$
d^2c_1({\cal L})=-2d\,\phi^*\omega
+\sum_{X\cup Y=S}\sum_{k=1}^d k^2{[}D_{d-k,k}^{X,Y}{]}.
\eqno\eq
$$
$D_{d-k,k}^{X,Y}$ is the locus of configurations
such that the point $x$ and the points marked by $X$ belong
to the branch of degree $d-k$, while the points marked by $Y$
belong to the branch of degree $k$.
This yields the recursion relation of intersection numbers
$$
\eqalign{
d^2\langle \sigma_n(\phi_{\alpha})\prod_{i\in S}& \sigma_{n_i}(\phi_{\alpha_i})
\rangle_{0,d}
=-2dn\langle \sigma_{n-1}(\phi_{\alpha+1})
\prod_{i\in S}\sigma_{n_i}(\phi_{\alpha_i})\rangle_{0,d} \cr
&+\!\!\!\!\sum_{X\cup Y=S}\sum_{k=1}^d k^2 n
\langle \sigma_{n-1}(\phi_{\alpha})\prod_{i\in X}\sigma_{n_i}(\phi_{\alpha_i})
\phi_{\beta}\rangle_{0,d-k}
\langle \phi^{\beta}\prod_{j\in Y}\sigma_{n_j}(\phi_{\alpha_j})
\rangle_{0,k}  \cr}
\eqno\eq
$$
which leads to \toprechori.

\endpage

\refout

\end